# Design of a 5G Multimedia Broadcast Application Function Supporting Adaptive Error Recovery

Carlos M. Lentisco, Luis Bellido, Andrés Cárdenas, Ricardo Flores Moyano, and David Fernández

*Abstract*—The demand for mobile multimedia streaming services has been steadily growing in recent years. Mobile multimedia broadcasting addresses the shortage of radio resources but introduces a network error recovery problem. Retransmitting multimedia segments that are not correctly broadcast can cause service disruptions and increased service latency, affecting the quality of experience perceived by end users. With the advent of networking paradigms based on virtualization technologies, mobile networks have been enabled with more flexibility and agility to deploy innovative services that improve the utilization of available network resources. This paper discusses how mobile multimedia broadcast services can be designed to prevent service degradation by using the computing capabilities provided by multiaccess edge computing (MEC) platforms in the context of a 5G network architecture. An experimental platform has been developed to evaluate the feasibility of a MEC application to provide adaptive error recovery for multimedia broadcast services. The results of the experiments carried out show that the proposal provides a flexible mechanism that can be deployed at the network edge to lower the impact of transmission errors on latency and service disruptions.

## I. Introduction

MOBILE data traffic is expected to increase sevenfold between 2017 and 2022 [1]. To satisfy this growing demand, an increase in network capacity is not enough; an improved utilization of available network resources is required. In long-term evolution (LTE) networks, this has been addressed by the use of broadcasting, as defined by the evolved Multimedia Broadcast/Multicast Service (eMBMS) [2] solution. Using eMBMS, the multimedia stream can be broadcast as a sequence of different files, called segments, following the dynamic adaptive streaming over HTTP (DASH) standard [3]. Segments that are lost due to broadcast transmission errors are individually recovered by the multimedia players using HTTP. This segment error recovery process adds a delay that has an impact on service continuity.

While 5G multimedia broadcast services have not yet been standardized [4], [5], their new design brings new and interesting opportunities to address the problems of error recovery in multimedia broadcast transmissions. In particular, 5G is expected to facilitate the deployment of multiaccess edge computing (MEC) [6] solutions, bringing the capabilities of cloud computing, traditionally available in centralized datacenters, to the edge of the network. Thus, it will be possible to deploy a multimedia broadcast application function at the edge of the network, acting as a cache, to reduce the delay associated with the segment error recovery process.

However, replicating the multimedia content at the edge of the network might not be enough. In [7], we showed how to avoid service disruptions in eMBMS using an adaptive error recovery mechanism that takes into account the bandwidth available in the unicast channel. In 5G, the integration of software-defined networking (SDN) [8] and network function virtualization (NFV) [9] solutions will make it possible to integrate this solution without incurring the costs associated with developing and implementing a new functionality in proprietary hardware.

The state-of-the-art shows some proposals defining how the LTE eMBMS service can be adapted to the 5G network architecture. However, existing proposals in this area do not describe how

This work was supported by the Spanish Ministry of Economy and Competitiveness and the Spanish Ministry of Science and Innovation in the context of ECTICS project under Grant PID2019–105257RB-C21 and Go2Edge under Grant RED2018–102585-T.

Carlos M. Lentisco, Luis Bellido, Andrés Cárdenas, and David Fernández are with the Departamento de Ingeniería de Sistemas Telemáticos, ETSI Telecomunicación, Universidad Politécnica de Madrid, Madrid 28040, Spain (e-mail: c.lentisco@upm.es)

Ricardo Flores Moyano is with Department of Computer Science Engineering, Universidad San Francisco de Quito, Quito 170901, Ecuador (e-mail: rflores@usfq.edu.ec)

MEC capabilities can be used to improve 5G multimedia broadcast services. While the integration between MEC and 5G is still an open issue, our proposal defines how a multimedia broadcast application function can be deployed at a MEC location, and how it is integrated into the 5G architecture.

Thus, this paper presents the design of a 5G multimedia broadcast application function integrating an adaptive error recovery mechanism. The paper explains how the multimedia broadcast application function is integrated into a novel 5G broadcast architecture. A 5G broadcast session start procedure is proposed in this paper to explain the relationships between all 5G broadcast network functions. The feasibility of implementing the multimedia broadcast application function is evaluated using an experimental platform. The results obtained using the experimental platform show how it is possible to apply adaptive error recovery mechanisms to reduce the impact of service disruptions and latency for multimedia broadcasting over a 5G network.

The rest of the paper is organized as follows: Section II reviews related work on mobile network architectures based on virtualization and mobile multimedia broadcast services. Standard mobile network architectures are presented in Section III. Section IV details how it is possible to support 5G multimedia broadcast services by combining NFV and SDN solutions. A MEC application function that can be integrated into the 5G network to avoid service degradation is specified in Section V. Section VI provides a description of the experimental platform that has been developed to evaluate the feasibility of implementing the MEC application function over standard NFV platforms. Finally, Section VII provides the conclusions of this work.

## II. BACKGROUND AND RELATED WORK

This section reviews related work on two different areas: design proposals for mobile broadcast architectures based on virtualization and proposals to improve service latency and stalling in mobile multimedia broadcast services.

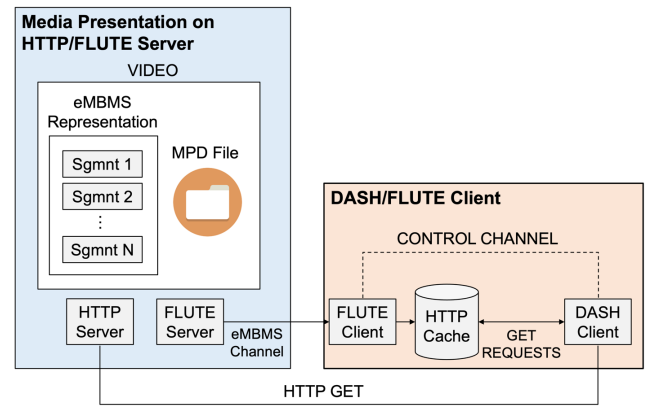

Fig. 1. Hybrid FLUTE/DASH architecture. This figure shows the components of the architecture supporting DASH in eMBMS.

### A. Mobile Broadcast Networks based on Virtualization

Standardization efforts [10] have been initiated by the 3GPP to define a 5G broadcast architecture based on virtualization technologies. Some proposals define how eMBMS services can be rearchitected by adopting SDN and NFV solutions.

Do et al. [11] proposed redesigning the logical entities of eMBMS as SDN applications. Functions performed by the eMBMS entities are regrouped depending on their scope. For example, all the broadcast session-related functions, which in LTE are distributed in several eMBMS logical entities, would be merged into an SDN application. This work, however, does not address how these SDN applications can be integrated into the 5G architectural framework.

Chantre and Fonseca [12], [13] defined how the eMBMS architecture can be virtualized from the perspective of NFV. They proposed migrating all the eMBMS functions, as they are defined for LTE, to a cloud computing platform. However, the Control and User Plane Separation (CUPS) [14] principle, which is one of the main pillars of the 5G network design, is not considered. CUPS is based on the idea of decoupling the control and user planes of mobile core network entities. This allows the integration of SDN solutions [15] to manage data plane traffic more efficiently.

Tran et al. [4] analyzed different alternatives to adapt the functional behavior of the eMBMS entities to the 5G network architecture. Our proposal follows a similar approach but also addresses how 5G multimedia broadcast services can be deployed in an NFV/SDN environment.

### B. Mobile Multimedia Broadcast Services

The 3GPP specification TS 26.346 [2] defines two delivery methods for live multimedia broadcast streaming services. The first method, named the *streaming delivery method,* uses the real-time protocol (RTP) to transport multimedia data packets over UDP. The second method, named the *download delivery method,* uses file delivery over unidirectional transport (FLUTE) [16] as the transport protocol. This paper focuses on the *download delivery method.*

In the *download delivery method,* multimedia content is encoded following the DASH standard. The DASH standard defines how a multimedia stream can be split into segments of the same duration. Segments belong to different representations, i.e., alternatives of the same content that are generated using different bitrates and resolutions. DASH players can adapt to the changing bandwidth conditions by switching between the available representations, which are defined in a metadata file named the Media Presentation Description (MPD). DASH was originally defined to be used over HTTP but 3GPP adapted its use for broadcast transmissions. Fig. 1 shows the architecture defined by the 3GPP for DASH streaming over eMBMS. Since the eMBMS channel is unidirectional, only one multimedia representation (eMBMS representation) is generated from the input stream. Multimedia segments are encapsulated as files that are sent using the FLUTE protocol, which works over UDP.

The delivery of files over eMBMS can be protected by applying application layer forward error correction (AL-FEC) techniques [17]. During the AL-FEC encoding process, redundancy packets are added to the multimedia data so that a client can decode a segment if it has received enough packets, either redundancy packets or multimedia data packets. Thus, the segment error rate depends on the packet error rate. Segments correctly broadcast are stored in a local cache that is located inside the user terminal. The DASH player can fetch these segments directly from the cache. Segments lost over the eMBMS channel (not decoded by the AL-FEC) are recovered using HTTP over a unicast channel.

One of the main research topics for DASH streaming is how to provide a seamless playback with a minimum delay [18]-[21]. The lower bound of delay for a DASH streaming service is the segment duration, since "a segment is a complete and discrete unit that must be made available in its entirety" [3]. In a mobile multimedia broadcast service, buffer starvation can occur if the broadcast transmission fails to deliver the multimedia segments. However, avoiding service disruptions is possible by adequately dimensioning the initial buffer [22].

With the advent of MEC technologies, current research is now focused on analyzing how multimedia streaming services can take advantage of the computing capabilities provided by MEC platforms. In this regard, three main topics have been identified: multimedia content caching [23], video transcoding [24], and multimedia adaptive streaming based on server and network-assisted DASH (SAND) [25], [26]. The proposal presented in this paper is based on a previous work [7] in which we showed how to avoid service disruptions in LTE broadcast services without overdimensioning the buffer of the player by using an adaptive error recovery mechanism.

In this paper, it is explained how the adaptive error recovery mechanism can be easily deployed as part of a 5G multimedia broadcast application function. Additionally, the paper also describes how this 5G multimedia broadcast application can be integrated into a novel 5G multimedia broadcast architecture.

### III. MOBILE NETWORK ARCHITECTURES

First, this section reviews the LTE network architecture, focusing on the logical entities that have been specifically designed for supporting eMBMS services. Then, the section provides a description of the 5G network architecture.

### A. LTE Network Architecture

Fig. 2 shows both the Evolved Universal Terrestrial Radio Access Network (E-UTRAN) and the Evolved Packet Core (EPC). The E-UTRAN is composed of a set of base stations. In LTE, a base station, also known as evolved NodeB (eNB), transmits the radio signal and controls the radio interface. In contrast, the EPC is composed of several functional entities specifically designed to support mobile multimedia broadcast services.

The Broadcast-Multicast Service Centre (BM-SC) is the entry point for content provider transmissions. To start a multimedia broadcast, the content provider must define different service and session parameters using the xMB-C reference point. Once the service is defined, data are sent over the xMB-U reference point from the content provider to the BM-SC and then over the eMBMS channel to the multicast receivers.

The BM-SC contains several software components that are used to broadcast multimedia content over LTE. It includes a FLUTE server [16], an AL-FEC encoder [17], and a web server. FLUTE is used to send the multimedia segments over the eMBMS channel. An AL-FEC encoder is typically used to add redundant data to the transmission that can then be used by the receiver to correct transmission errors. However, even an AL-FEC encoded segment can be lost during its broadcast transmission. In that case, segments can be recovered using HTTP directly from the BM-SC web server.

The Multimedia Broadcast Multicast Service Gateway (MBMS GW) is an eMBMS entity that uses IP multicast to forward eMBMS data to all the base stations that belong to the same area. In addition, it performs session control signaling (start/update/stop session) toward the E-UTRAN through the mobility management entity (MME), the main control entity for LTE.

Finally, the 3GPP defines a control plane entity called the multicell/multicast coordination entity (MCE) that performs two main tasks. It determines the amount of radio resources that will be allocated to the eMBMS service, and it sets up radio transmission parameters such as the Modulation and Coding Scheme (MCS) [17]. Both the MCS and the allocation of radio resources determine the available bandwidth in an eMBMS channel. Bandwidth in a multicast channel is allocated statically, so these parameters, which are specified during the establishment of the session, cannot be changed on demand.

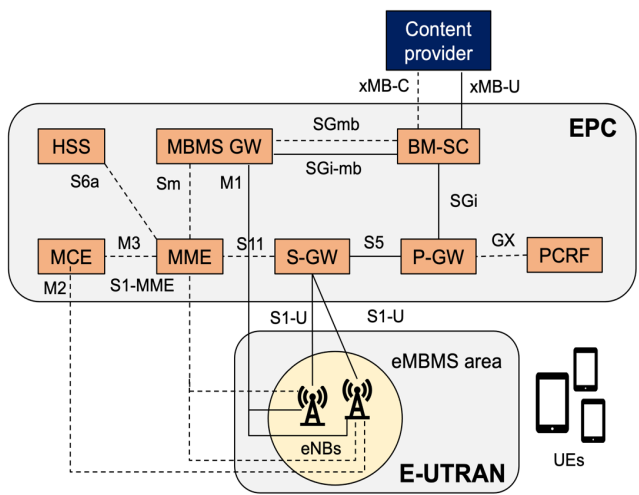

Fig. 2. Network architecture model for LTE. This figure shows the logical entities designed for supporting eMBMS: BM-SC, MBMS GW, and MCE.

Fig. 2 also shows other functional entities supporting unicast services. The Serving Gateway (S-GW) and the Packet Data Network Gateway (P-GW) are in charge of forwarding unicast data in the mobile network. Regarding eMBMS services, Fig. 2 shows that the SGi interface makes it possible to establish an HTTP connection to the web server located at the BM-SC to recover multimedia segments that have been lost over eMBMS.

### B. 5G Network Architecture

In the 5G network architecture, control and user plane functions of the 5G core network entities are decoupled, following the CUPS principle. Fig. 3 shows a partial view of the 5G core network architecture [27], [28]. Only the functional entities more closely related to the proposal of a new broadcast multimedia architecture are depicted. These functional entities are described below.

The access and mobility management function (AMF) controls the registration and deregistration of user terminals in the network. It performs mobility management functions for tracking the user terminal location in the network and functions that support handover processes.

The session management function (SMF) is responsible for the management of session-related functions. The SMF programs data plane network functions for establishing the route paths that allow clients to access different services. It also hosts a DHCP server that assigns an IP address to the terminal client when a new session is established. In 5G networks, all user plane functions are integrated into the user plane function (UPF) entity, which works similarly to an SDN switch. In that sense, the SMF can be seen as an SDN controller that programs

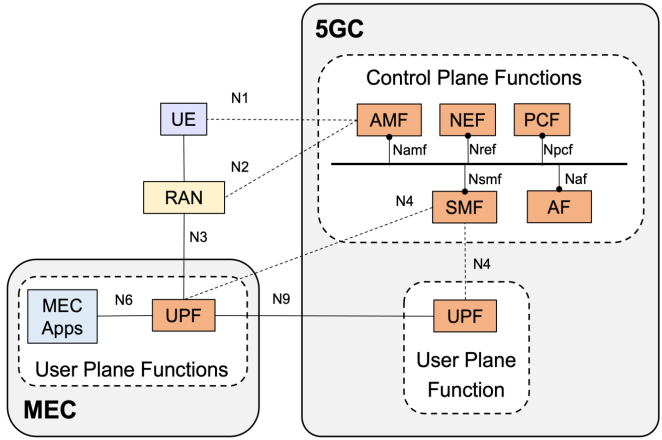

Fig. 3. MEC in 5G network architecture. This figure shows a partial view of the network architecture model for 5G, in which some user plane functions can be moved to a MEC platform.

the network of UPFs to provide communication between terminal clients and services.

The policy control function (PCF) encompasses different charging and policy control functions related to user mobility and the management of service data flows [29]. For example, it manages policies that are used by terminal clients to determine how the outgoing traffic must be routed and policies specifying how to apply a specific QoS treatment to a particular service data flow.

The 3GPP defines the application function (AF) as a functional entity that can provide different kinds of services. To support the provision of a service, an AF can influence traffic routing or quality of service by interacting with 5G core network entities. A mobile network operator can give permissions to third-party application functions to use the services provided by the 5G core network entities. In this case, the AF accesses the 5G services through a secure interface provided by the network exposure function (NEF).

In this paper, an application function supporting adaptive error recovery for multimedia broadcast services is proposed. Section IV. B shows how the proposed application function interacts with the 5G core during the establishment of a 5G broadcast session.

Finally, Fig. 3 shows how MEC platforms can be integrated into 5G networks. A local UPF is colocated with the MEC platform so user traffic can be redirected to MEC applications deployed at the edge. When the mobile terminal is registered in the network, this local UPF is chosen as the packet switching node associated with the packet data session. UPF selection depends on the user terminal location and on AF information, and it can happen during the establishment of a data session or when an AF sends a request to modify routing paths.

## IV. Supporting 5G multimedia broadcast services

This section is divided into two parts. Section IV. A describes a novel 5G multimedia broadcast architecture that includes a multimedia broadcast application function supporting adaptive error recovery. Section IV. B describes a new 5G broadcast session start procedure.

The design of the 5G multimedia broadcast architecture is based on the CUPS principles and the adoption of new network paradigms such as NFV, SDN, and MEC. Control plane functions usually require a low latency to process signaling messages. Data plane functions, on the contrary, demand more bandwidth to process data traffic. CUPS allows scaling both planes independently to satisfy these requirements in a more efficient manner. In 5G, all the data plane functions have been integrated into a single network entity, the UPF. Control plane functions have been modularized according to their scope (session, mobility, or authentication).

Regarding the adoption of new network paradigms, ETSI notes [30] that MEC applications can be deployed as VNFs controlled by an NFV management and network orchestration framework (NFV-MANO) [31]. MANO is mainly composed of three elements: a virtualized infrastructure manager (VIM), an NFV orchestrator (NFVO), and a VNF manager (VNFM). The network, storage, and computing resources available at the NFV Infrastructure (NFVI) are controlled by the VIM. The lifecycle of network services and VNFs is managed by the NFVO and the VNFM.

### A. 5G Broadcast Architecture Supporting Multimedia Streaming Services

Fig. 4 shows our design proposal for a 5G multimedia broadcast architecture. The figure shows the radio access network, a MEC platform, and the 5G core (5GC) network.

The radio access network in 5G is based on the idea of virtualizing part of the functionality performed by base stations on a cloud-based platform. Base stations integrate two elements: a radio remote header (RRH) and a base band unit (BBU). The RRH is the component that transmits the radio signal. The BBU processes the signal and performs other functions, such as radio resource scheduling or radio link control. Cloud-RAN (C-RAN) [32] proposes moving the BBUs to the cloud to be virtualized. This deployment, as shown in Fig. 4, can be done at the edge of software-defined central offices [33].

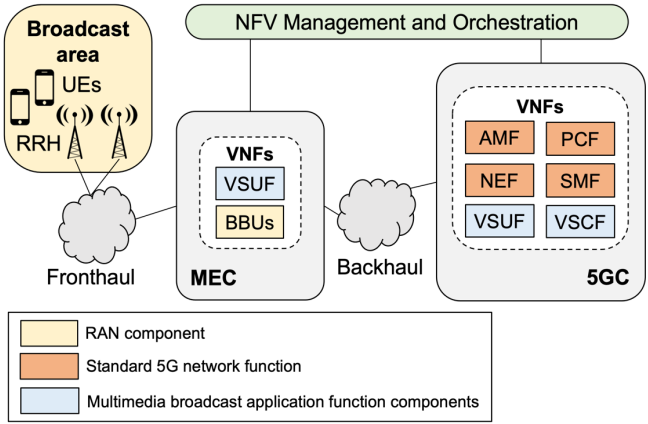

Fig. 4. 5G broadcast architecture proposal. The figure shows how a Video Streaming User plane Function (VSUF) can be deployed both at the MEC facilities or at the 5G core network.

A 5G broadcast area is composed of several base stations that transmit the same content at the same time. All the base stations that belong to the same broadcast area are connected to the same MEC through the fronthaul network. A broadcast transmission can cover several broadcast areas, so 5G multimedia broadcast services can be provided from different MEC locations at the same time.

Thus, MEC applications and BBUs would be deployed by an NFV-MANO platform as VNFs at the network edge, while the control plane functions identified in the previous section (SMF, AMF, PCF and NEF) would be deployed at the 5GC, as recommended by the 3GPP standards and other ETSI reports [27], [34]. The SMF acts as an SDN controller that programs UPFs and includes a DHCP function and a mechanism used to collect charging data.

The video streaming user plane function (VSUF) is proposed to support DASH streaming services in 5G networks. VSUF is able to deliver DASH-formatted content using HTTP over a unicast channel or using a file delivery protocol such as FLUTE over the 5G broadcast channel, as explained in Section II.B. Multimedia functions performed by VSUF are controlled by the video streaming control plane function (VSCF).

VSUF includes multimedia functions that are used to store and prepare multimedia content that will be transmitted over the 5G network. Multimedia functions such as a cache server, a video transcoder, or a SAND proxy [26] can be included in VSUF. VSUF also supports 5G multimedia broadcast services, so it includes functions specifically used in broadcasting scenarios, such as an AL-FEC encoder/decoder or a FLUTE server. Thus, VSUF implements the user plane of the BM-SC LTE entity.

The VSCF, as a 3GPP compliant application function, will make decisions on traffic routing, programming the network of UPFs by using the network capabilities exposed by the SMF. IP multicast data packets sent by VSUF will need to be forwarded by UPFs to the BBUs that belong to the mobile broadcast area. BBUs will send these packets over a common channel to the multicast receivers. When a multimedia segment is lost over the multicast channel, the client will need to retrieve it using HTTP. Consequently, the network of UPFs will also need to be programmed so that the HTTP request sent by the client can be delivered over a unicast route.

As Fig. 4 shows, VSUF can be deployed at the central 5G core or at a MEC platform as a MEC application. Thanks to the flexibility provided by NFV technologies, its behavior can be easily modified to incorporate new multimedia streaming mechanisms specifically designed to improve key parameters of the service. Section V explains how a multimedia adaptive error recovery mechanism can be used to reduce the service latency of 5G multimedia broadcast services.

### B. 5G Broadcast Session Start Procedure

5G standards have not yet specified the procedures for managing 5G broadcast sessions. Therefore, this section describes a proposal for a 5G Multimedia Broadcast Service (5GMBS) session start procedure (Fig. 5) that combines three existing standard procedures: the MBMS service creation procedure [2], the MBMS session start procedure [35], and the Protocol Data Unit (PDU) session establishment procedure [28].

The 5GPPP 5G-Xcast project made a similar proposal for a network resource allocation procedure [10]. Differences and similarities between both proposals will be remarked upon in the description provided below. However, the main objective of this subsection is to explain the relation between VSUF, VSCF, and other 5G network functions.

The 5GMBS session start procedure is triggered by the content provider, which sends a request to VSCF (1) to specify several session and service parameters that define how multimedia content will be broadcast over 5G. The parameters defined in the 3GPP LTE standard [2] can also be used for 5G broadcasting. These parameters are used, among other things, to specify a maximum video bitrate or the geographical area where multimedia content is broadcast.

The multimedia broadcast application function could be configured at this point, as proposed by the 5G-Xcast project. However, broadcast sessions cannot be established with success if network resources have not been previously reserved at the RAN and the 5GC. The 5GMBS session start procedure waits until this configuration is done to avoid unnecessarily configuring the VSUF.

Once the VSCF has received the content provider request, session and service parameters are transmitted to the AMF assigned to the terminals subscribed to the service (2). An AMF sends this information to RAN (3) using the N2 reference point (Fig. 3) so that radio resources can be allocated to provide an adequate level of quality of service (4.a).

Once the AMF has received the 5G start request message (2), an SMF is selected to manage the PDU session. A PDU session provides a connection between a terminal user and a data network, for example, the internet or a private corporate network. PDU sessions are identified univocally and, in general, are established on demand by terminals. Considering that broadcast data are sent over a common channel to all the multicast receivers, only one PDU session is needed. The geographical area where multimedia content is broadcast can be as wide as several MEC platforms, i.e., different 5G broadcast areas. In this case, a PDU session for each area needs to be created. This kind of PDU session, hereafter referred to as a multicast PDU session, may be created at the request of the multimedia content provider. The multicast receiver joins this multicast PDU session, as indicated in the 23.502 3GPP technical specification [35], i.e., sending the AMF a message that expresses its intention to join an existing PDU session [28].

The SMF selection (4.b) depends on the geographical area where the multimedia is going to be broadcast. The selected SMF will program the network of UPFs (9.a) to deliver the multimedia content from the origin server to the base stations that belong to the broadcast area. As proposed by the 5G-Xcast project, the selected SMF will receive broadcast area information and the Data Network Name (DNN).

At some point, the AMF will receive a response from the RAN, indicating that the radio resource allocation process has concluded (5). With the SMF already selected, a context for the multicast PDU session is then created (6). This context would be very similar to the MBMS bearer context defined by the 3GPP for an LTE broadcast service [2]. The PDU session ID, the DNN, the geographical area, or an IP multicast address would be included in this context. In LTE, the IP multicast address is allocated by the MBMS GW. We propose that this task be performed by the SMF, since it is already responsible for allocating IP addresses to clients when the establishment of a new PDU session has been requested.

The 5GMBS session start procedure continues with UPF and PCF selection (7), which is also based on broadcast area information. The selection of the PCF can be followed by an SMF request to obtain the Policy and Charging Control (PCC) rules that have been defined for the multicast PDU session. A PCC rule consists of a set of parameters that are used for detecting, charging, and applying QoS policies to a service data flow.

When a content provider requests service creation, several parameters related to the QoS can be provided to the VSCF application function. In LTE, the following two parameters are considered [2]: the maximum video bitrate and the maximum delay. The VSCF, as an application function, can provide this information to the PCF (8.a) [29], which creates a new PCC rule for the multicast PDU session. Then, the SMF obtains this PCC rule (8.b), programming the data plane so that the QoS requirements specified by the content provider can be guaranteed. The relation between the SMF, the PCF and the multimedia broadcast application function has not been explained in previous work.

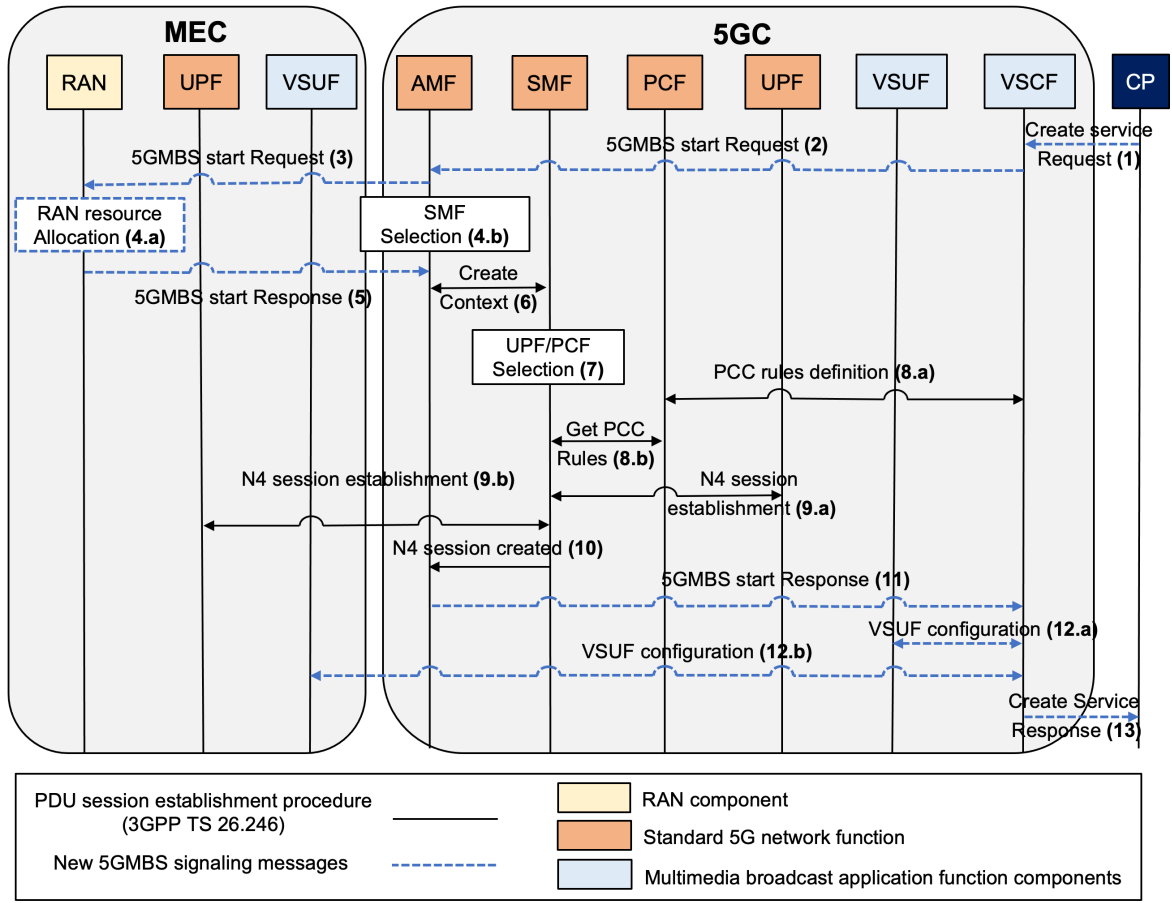

Fig. 5. 5G Broadcast Multimedia Service (5GMBS) session start procedure. This figure shows the signaling messages required to initiate a 5G multimedia broadcast service.

The SMF programs, using the N4 interface, all the UPFs connecting user terminals to the serving VSUF instances, regardless of their location at either the central 5GC (9.a) or, as we propose, at the MEC facilities (9.b). Then, the SMF sends a message to the AMF as a notification that the multicast data flow has been installed in the data network (10). The AMF, in turn, sends a response to the VSCF that requested the creation of a 5G broadcast session (11).

At this point, if all the above processes have been completed successfully, the VSCF configures the VSUF services (12.a, 12.b) according to the content provider specifications. Finally, the VSCF reports the result of the 5G broadcast session establishment procedure (12) to the content provider. If the result is positive, the content provider can inject multimedia content into the network for its distribution, as will be explained below.

## V. 5G MULTIMEDIA BROADCAST SERVICES BASED ON DASH

The proposed multimedia broadcast application function, i.e., VSUF, implements an adaptive error recovery mechanism that can be used to reduce the service latency of 5G multimedia broadcast services. This section explains how VSUF acts as a SAND adaptive recovery proxy (SARP) that provides adaptability to the recovery of segments that have been lost during broadcast transmission.

Initially, and as Fig. 6 shows, the 5GMBS session start procedure described in Section IV. B is carried out under the request of the content provider (1). Then, Fig. 6 shows, as an example, that multimedia segments are sent from the content provider to a VSUF deployed at the MEC using a push mode (2). That is, the content provider sends multimedia segments to VSUF as soon as they are available.

Multimedia content is prepared to be broadcast (3) so multimedia segments can be sent from VSUF (4)

to all the BBUs that belong to the same broadcast area. Segments that cannot be recovered by the AL-FEC (5) will be requested by the DASH player by sending an HTTP GET request to VSUF (6), as explained in Section II.B.

At this point, VSUF accesses the bandwidth information from the BBU that will be in charge of forwarding the multimedia segment over the unicast channel. Bandwidth information is obtained in near real-time from the Radio Network Information Service (RNIS) [36] of the MEC platform. By using the RNIS REST API, VSUF queries the amount of bandwidth allocated by the BBU radio resource scheduling algorithm to the unicast channel of the specific terminal sending the HTTP GET request (7).

Once VSUF has retrieved the bandwidth information, an adaptation algorithm selects the multimedia representation (RepQ) that best fits the available bandwidth. This can be calculated as follows:

$$\text{RepQ} = \begin{array}{c} \arg\max B_k \\ k \\ \text{subject to } B_k < BW_{unicast} \end{array} \qquad (1)$$

where $B_k$ is the bitrate of the kth representation and $BW_{unicast}$ is the available unicast bandwidth. Finally,

VSUF sends the selected representation to the client (8). If a specific multimedia representation segment is not stored in the cache located at VSUF, VSUF can either request it to the content provider or generate it on-the-fly using a video transcoder.

## VI. RESULTS

First, this section describes an experimental platform that has been developed to evaluate the feasibility of implementing VSUF over standard NFV platforms. Then, the proposal is evaluated by obtaining measurements of the live latency, defined as the difference between the time at which a live event occurs and the time at which the live event is played on a receiving terminal. In DASH streaming, live latency is affected by the number of segments that multimedia players store in their buffer before the multimedia playback can start. The *minBufferTime* attribute of the MPD file [3] defines this number. In this paper, the low latency mode of a reference DASH player implementation is used, obtaining an initial buffering delay of one and a half segments.

### A. 5G Multimedia Broadcast Service Analysis Platform

The experimental platform was developed using virtual networks over linuX (VNX) [37]. VNX is a virtualization tool that allows building virtual network testbeds based on Linux Containers (LXC) and kernel-based virtual machines (KVMs) that can be configured as servers, routers, or clients. The use of Linux bridges and Open vSwitch (OVS) provides connectivity at level 2. The validation scenario that has been developed for analyzing 5G multimedia broadcast services uses VNX to emulate the mobile terminal client, the fronthaul network, and the RRH providing network access to the mobile terminal client.

The mobile terminal client consists of a multimedia player and a local cache that stores the multimedia segments that are broadcast over 5G. Segments that have been lost during its broadcast transmission are recovered using HTTP from the VSUF deployed at the MEC. This functionality has been implemented by using a standard Squid server that has been installed in the LXC container that acts as the local cache. The reference implementation of a DASH client developed by the DASH industry forum (dash.js) has been used to access the multimedia content. Multimedia content was prepared for transmission using the FFmpeg[1] and MP4Box[2] video encoding tools.

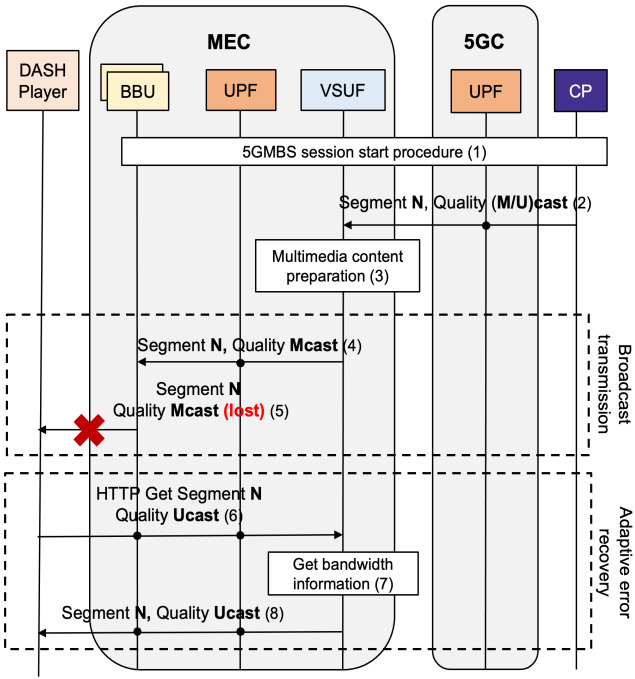

Fig. 6. Adaptive multimedia recovery based on server and network assisted strategies. This figure shows the VSUF application providing a representation for error recovery based on the available bandwidth.

| Parameter | Value |
|---|---|
| Length of video | 600 s |
| $T_{smgnt}$ | 0.5 s |
| $BW_{unicast}$ | 2/2.5/3/3.5/4 Mbps |
| Video bitrate | 3/6 Mbps |
| Video resolution | 1080p |
| Initial buffering delay | 0.7 s |
| Segments recovered by HTTP | 10% |

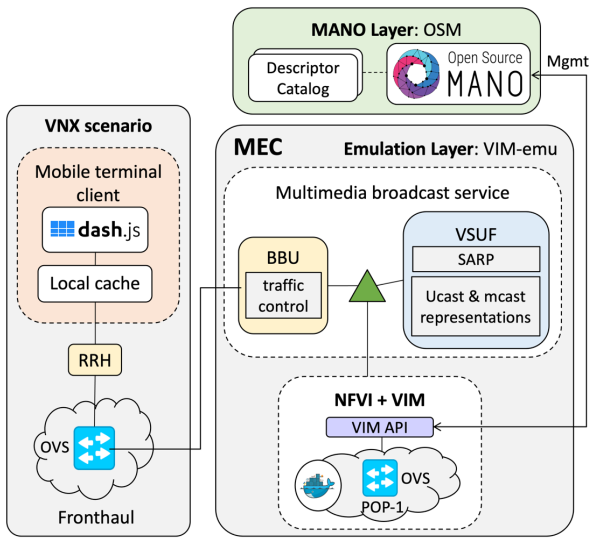

Fig. 7. Block diagram of the experimental platform used to evaluate the performance of the VSUF.

The RRH has been emulated using an LXC container, but for the time being, it only forwards traffic from the mobile terminal client to the BBU deployed at the MEC. The fronthaul network has been emulated using an OVS switch that connects the RRH to a software-defined central office that hosts the VNFs and MEC applications deployed at the network edge.

The management and orchestration platform used at the MEC is Open Source MANO (OSM) [38], which acts as both the NFV orchestrator and the VNF manager, controlling the NFVI resources that are emulated using Vim-emu[3]. Vim-emu deploys the VNFs that compose a network service packaged as Docker containers. As shown in Fig. 7, all the VNFs composing the network service are internally connected to an OVS switch provided by Vim-emu and hosted inside one of the Docker containers that implement the Vim-emu functionality.

The multimedia broadcast network service is composed of two VNFs: a BBU and VSUF. In this simplified version of the scenario, the BBU and VSUF are interconnected through an OVS switch based on traditional switching methods. The BBU controls the bandwidth that is allocated to the mobile terminal client, a feature that has been emulated by using the Linux traffic control (*tc*) tool. Both the BBU and VSUF have been defined by means of descriptors written in YAML that have been on-boarded into the OSM catalog.

### B. Performance Evaluation of a SAND Proxy for Adaptive Error Recovery

Table I shows the service parameters used to evaluate the SARP included in the VSUF. The Big Buck Bunny[4] video has been encoded considering a segment duration ($T_{sgmnt}$) of 0.5 s, which allows a reduction of service latency and reaching high levels of quality of experience [39]. To limit the initial buffering delay, the dash.js player low latency mode is used. This low latency mode provides an initial buffering delay of approximately 0.7 s, as measured in the experiments carried out. In this evaluation, it is assumed that ten percent of the segments need to be recovered using HTTP and that segment losses follow a uniform distribution.

Multimedia segment transmission errors are simulated in the experimental platform as follows: all the segments that compose the multimedia presentation are prestored in the mobile terminal client local cache (Fig. 7). Segments that are considered lost are identified in a plain text file (hereinafter referred to as *error file*). When the dash.js player requests a segment that is included in the *error file*, the local cache forwards the segment request to the SARP to recover it using HTTP. For the experiments, ten different *error files* were generated using the uniform distribution function.

Table I also shows that two multimedia representations are available, one with a bitrate of 6 Mbps and another with a bitrate of 3 Mbps, both with a resolution of 1080p and generated using a variable bitrate (VBR) coding method. The 6 Mbps representation is used for broadcasting. The alternative 3 Mbps representation will be selected for unicast recovery when the available unicast bandwidth is below the bitrate of the broadcast representation. This alternative representation will only be selected when the SARP is used.

Three different experiments have been carried out. The first two experiments analyze the impact of a low available unicast bandwidth on live latency when SARP is not used (Fig. 8) and when SARP is used (Fig. 9). For these two experiments, the unicast bandwidth is set to a constant value during the whole multimedia playback. The third experiment aims to evaluate the proposal in a more realistic environment by using bandwidth measurements obtained from a demonstration LTE network [40].

Table I shows the unicast bandwidths that were selected to carry out the first two experiments: 2, 2.5, 3, 3.5, and 4 Mbps. This set of values has been selected to analyze how the live latency evolves when the unicast bandwidth is higher than, equal to, or lower than the bitrate of the 3 Mbps alternative representation.

Fig. 8 shows the live latency that is measured by the dash.js player when SARP is not used. The results are obtained using the same *error file* for all the unicast bandwidths evaluated and considering that segment 120 (playback time 60 s) is the first segment that is lost in the multicast channel. Lost segments are requested with the default quality of 6 Mbps. Since these segments are retrieved from VSUF using HTTP over the unicast channel, the segment transmission delay can be longer than segment duration, leading to buffer starvation [22].

Fig. 8 shows the two effects of buffer starvation in the multimedia playback: stalling and an increase in live latency. The playback stalls several times during the experiment. For example, at the beginning of the multimedia playback, when the first lost segment is recovered from VSUF (at playback time 60 s), the buffer level of the DASH client is too short to prevent buffer starvation. Because of buffer starvation, the dash.js player will increase the size of the buffer to avoid future stalling, and therefore, live latency will be increased. As Fig. 8 shows, this initial increase in the size of the buffer may be insufficient to prevent more stalling. At the end of the experiment (playback time 300 s), live latency increased to 2.3, 2.47, 2.8 and 3.4 s for unicast bandwidths of 4, 3.5, 3 and 2.5 Mbps, respectively. Since 4 Mbps is a value greater than the bitrate of the alternative representation, theoretically, playback should not suffer from stalling. In this case, several factors cause the live latency to increase. The Big Buck Bunny video has been encoded using the VBR method, so the size of the DASH segments varies within the multimedia presentation. On the other hand, the *tc* tool has been configured to work as a hierarchical token bucket, which makes the available unicast bandwidth vary around the value of 4 Mbps.

When the unicast bandwidth is too low, a long stalling event can occur. For instance, Fig. 7 shows that stalling caused the live latency to increase 2 s from a playback time of 126 s to 133 s for a unicast bandwidth of 2 Mbps. Live latency only increased 0.28 s during the same period for a unicast bandwidth of 4 Mbps. For long stalling events, rebuffering causes live latency to exceed the levels that are adequate for a live streaming service. However, after such events, the dash.js player accelerates the playback rate to reduce live latency. Fig. 8 shows that this happened for a unicast bandwidth of 2 Mbps. For instance, live latency was reduced 2.7 s from playback time 133 s to 196 s.

Fig. 9 shows the live latency that is measured by the dash.js player when SARP is used. In this case, multimedia segments are broadcast with the default quality of 6 Mbps, and the segments that are lost in the multicast channel are retrieved from the SARP with the alternative quality of 3 Mbps. Ten different *error files* were used to obtain results with a confidence level of 90%.

Fig. 9 shows that as the unicast bandwidth increases, there is a lower increase in live latency. At the end of the experiment (playback time 300 s), live latency increased to 1.3, 1.4, 1.5, 1.7 and 2.1 s for unicast bandwidths of 4, 3.5, 3, 2.5 and 2 Mbps, respectively. Fig. 9 also shows that the duration of the stalling events is more limited when SARP is used. With a confidence level of 90%, the playback stalled, at most, 0.13 s between playback times of 73 s and 88 s for the worst-case scenario, i.e., considering a unicast bandwidth of 2 Mbps. Therefore, in this case, multimedia playback has not suffered from long stalling events, and therefore, the dash.js player has not tried to adapt the playback rate to reduce live latency.

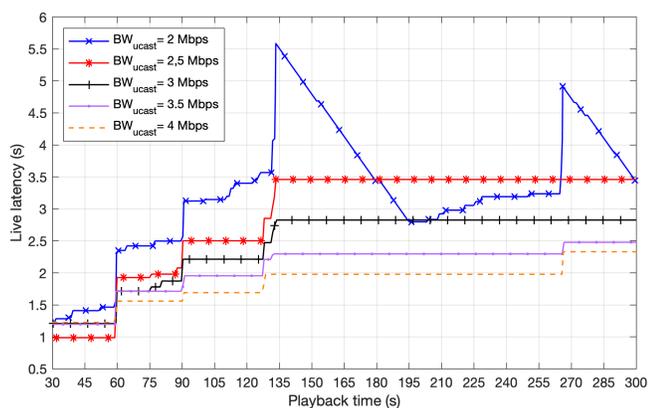

Fig. 8. Live latency evolution without using SARP. This figure shows how live latency evolves for different values of the unicast bandwidth ($BW_{ucast}$).

Finally, Fig. 10 shows how the SARP behaves in a scenario in which the unicast bandwidth is based on real LTE bandwidth measurements [40]. With a confidence level of 90%, the results show the live latency measured by the dash.js player for two different bandwidth profiles. *BW profile A* (Fig. 10c) presents low bandwidth fluctuations, while *BW profile B* (Fig. 10d) presents high bandwidth fluctuations. The mean of the bandwidth measurements is 3.6 Mbps for *BW profile A* and 2.6 Mbps for *BW profile B*. The variance is 0.82 and 2.49, respectively.

For *BW profile A*, Fig. 10a shows that if the SARP is not used, live latency increases at playback time 300 s to 2.6 s. Using the SARP, live latency can be reduced to 1.7 s. For *BW profile B*, Fig. 10b shows that the increase in live latency is mainly caused by two rebuffering events, which occur at playback times 50 s and 222 s. The dash.js player tries to adapt the playback rate to reduce live latency, but even if the SARP is used, the multimedia quality is severely affected by these two rebuffering events.

The performance evaluation was carried out using a uniform distribution function that models how the multimedia segments are lost in the eMBMS channel. The live latency measured during this evaluation would increase for segment loss distributions that cause several consecutive segments to be lost during the broadcast transmission, in those cases where the unicast bandwidth was lower than the bitrate of the lowest quality representation. However, the results obtained from the three experiments verify that by using SARP, it is possible to reduce the increment of live latency on the multimedia playback.

## VII. CONCLUSIONS AND FUTURE WORK

Mobile multimedia broadcast services are typically subject to multiple factors affecting service performance, such as latency or service disruptions. However, traditional mobile network architectures lack the flexibility needed to deploy new streaming mechanisms designed to prevent service degradation in an agile manner.

This paper proposes the design of a 5G multimedia broadcast architecture facilitating the deployment of multimedia functions at the network edge. Based on this design, an adaptive error recovery MEC application is defined to lower the impact of errors in latency and service disruptions for broadcast services. The proposal has been evaluated by developing an experimental platform based on a standard NFV platform to carry out several experiments that show promising results.

As a future step, experiments considering more realistic 5G bandwidth profiles will be carried out to explore how the proposed MEC application function behaves in different network scenarios. Further research will focus on the best strategies to encode multimedia content to be broadcast over 5G using DASH.

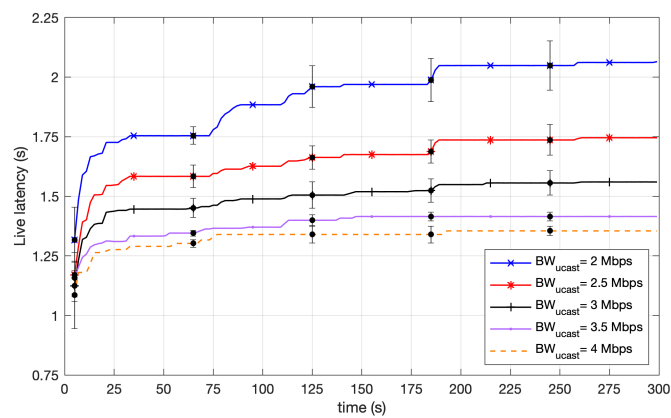

Fig. 9. Live latency evolution when using SARP. This figure shows how live latency evolves for different values of the unicast bandwidth ($BW_{ucast}$).

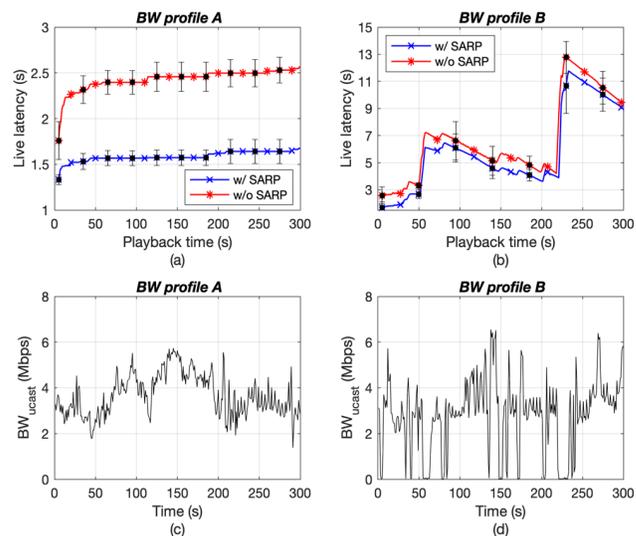

Fig. 10. Experiments carried out using real LTE bandwidth measurements. The figure shows how the live latency evolves when SARP is used (w/SARP) or not (w/o SARP), considering two different bandwidth profiles.

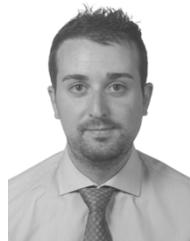

**Carlos M. Lentisco** received the M.S. and Ph.D. degrees in telecommunications engineering from the Universidad Politécnica de Madrid (UPM), Spain, in 2014 and 2019, respectively. He is currently a Teaching Assistant with UPM, specializing in the fields of computer networking, multimedia services, and the internet technologies. His current research interests include mobile broadcast services, virtualization, and software defined networking.

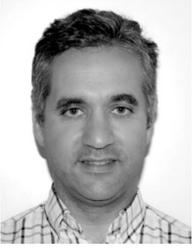 **Luis Bellido** received the M.S. and Ph.D. degrees in telecommunications engineering from the Universidad Politécnica de Madrid (UPM), Madrid, Spain, in 1994 and 2004, respectively. He is currently an Associate Professor at UPM, specializing in the fields of computer networking, internet technologies, and quality of service. His current research interests include mobile networks, multimedia applications, and virtualization

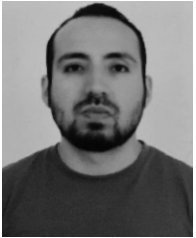 **Andrés Cárdenas** received a master's degree in network engineering and telematic services from the Universidad Politécnica de Madrid, Spain, where he is currently pursuing a Ph.D. degree with the Departamento de Ingeniería de Servicios Telemáticos (DIT). His research interests include NFV/SDN environments, network slicing, and 5G networks.

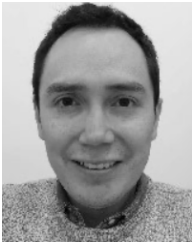 **Ricardo Flores Moyano** (Member, IEEE) received a degree in electronic engineering from Universidad Politécnica Salesiana (UPS) in 2006, and an M.S. and Ph.D. degrees in telematic systems engineering from the Universidad Politécnica de Madrid (UPM) in 2013 and 2018, respectively. He is a full-time professor with the Department of Computer Science Engineering, Universidad San Francisco de Quito (USFQ). His research interests include SDN, NFV, cloud networking, the future internet, and 5G networks.

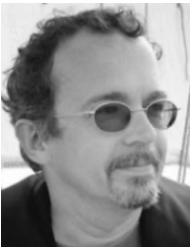 **David Fernández** received the M.S. degree in telecommunications engineering and a Ph.D. degree in telematic engineering from the Universidad Politécnica de Madrid (UPM), Spain, in 1988 and 1993, respectively. Since 1995, he has been an Associate Professor with the Department of Telematics Systems Engineering (DIT), UPM. His current research interests focus on software-defined networks, network virtualization, cloud computing datacenter technologies, and network security.